\documentclass[10pt,letterpaper,twocolumn]{article} %% two column, final layout
\usepackage{graphicx} 
\usepackage{ol2}
\usepackage[draft]{hyperref}
\usepackage{amsmath}

\begin{document}

\twocolumn[ %% activate for two-column option

\title{Measurement of transport mean free path of light in thin systems}

%% For REVTeX it is possible to automate superscript and e-mail callouts with the superscriptaddress option; see REVTeX4 documentation.

\author{Marco Leonetti,$^{1,*}$ Cefe L\'opez,$^1$ }

\address{
$^1$Instituto de Ciencia de Materiales de Madrid (CSIC) \\ and Unidad Asociada CSIC-UVigo, Cantoblanco \\ 28049 Madrid
Espa\~{n}a.
\\
$^*$Corresponding author: marco.leonetti@icmm.csic.es
%$^2$Editorial Services Department, Optical Society of America, \\ 2010 Massachusetts Avenue, NW,
%Washington, D.C. 20036, USA \\
%$^3$Currently with the Electronic Journals Department, Optical Society of America, \\ 2010 Massachusetts Avenue, NW,
%Washington, D.C. 20036, USA \\
%$^*$Corresponding author: xyx@osa.org
}

\begin{abstract} We extensively investigate in-plane light diffusion in systems with thickness larger than but comparable with the transport mean free path. By exploiting amplified spontaneous emission from dye molecules placed in the same holder of the sample, we obtain a directional probe beam precisely aligned to the sample plane. By comparing spatial intensity distribution of laterally leaking photons with predictions from random walk simulations, we extract accurate values of transport mean free path, opening the way to the investigation of a previously inaccessible kind of samples.\end{abstract}


] %% activate for two-column option

The study of light transport in turbid materials is an active field of research with many interdisciplinary applications in biomedicine \cite{Tromberg:97}, atmospheric study, cosmology \cite{Rad_Transf}, archaeology \cite{leonetti:101101} and also fundamental physics \cite{Barthelemy2008}. White, diffusive materials are composed of many disordered particles that perturb the propagation direction of light rays with respect to the straight trajectory. The usual strategy to measure the transport mean free path $\ell$, (that is the length after which a light ray loses the memory of the incoming direction) consists in comparing intensity outgoing from the sample with predictions of diffusion theory.  The most popular techniques  are  enhanced backscattering cone\cite{PhysRevLett.55.2692, PhysRevLett.55.2696} and total transmission measurements\cite{PhysRevB.46.14475, PhysRevA.78.023823}. The former requires a sample with a large scattering volume and the second needs many samples with variable volume. Probing in-plane properties of thin systems with this two approaches, has to face alignment difficulties and reduced signal to noise ratio \cite{P_sheng} due to losses.  $\ell$ may be extracted also by measuring spatial distribution of laterally leaking intensity \cite{Taniguchi:07, Johnson:08}, but up to now this approach has been limited to large samples with $\ell$ of the order of millimeter. Here we develop a measurement protocol that overcomes alignment difficulties, reduces unwanted scattering produced at interfaces and allows to measure $\ell$ in systems with thickness of the order of tens of micrometers.

\begin{figure}[h]
\includegraphics[width=8 cm]{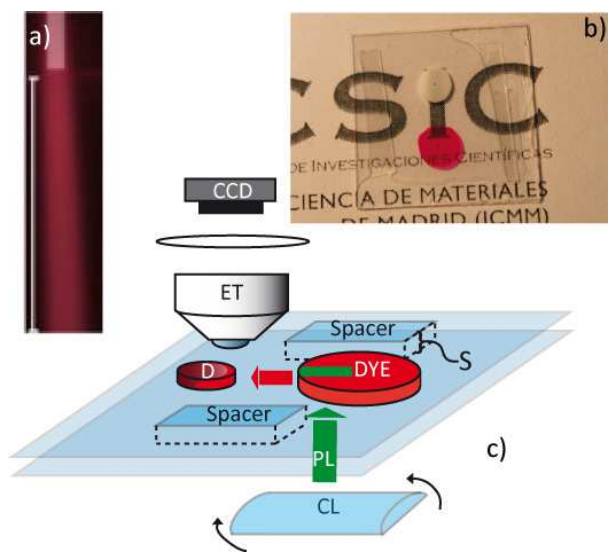}
\caption{\label{fig:setup}(Color Online) a) image of the light scattered by the ASE beam in the rhodamine solution; b) image pf two drops inside the sample holder. The shaded rectangle indicates the pumping area and white line is 870 $\mu$m long. c)Setup of the experiment. A stripe shaped beam from a nanosecond Nd:YAG pulsed laser (PL), shines on the rhodamine solution (DYE), producing pencil of light emitted at the edges of the inverted area. The amplified spontaneous emitted light impinges on the diffusive sample (D). Light leaked from the sides of the sample is collected on the image plane of a x10 enlarging telescope (ET) and collected by the CCD array of a camera.}
\end{figure}

We employ as a test diffusive system a water dispersion of latex particles (Duke Scientific size 520 nm). The sample holder (two microscopy coverglasses separated by plastic spacers of thickness $S$=100 $\mu$m) contains a (2 $\mu$l) drop of diffusive liquid distant about 2 millimeter from a drop of 3 $\mu$l drop of diethylene-glycol with 1\% volume of rhodamine B (see figure \ref{fig:setup}b).  As a directional light source we used the amplified spontaneous emission (ASE\cite{Siegman}) generated between the two coverglasses and in the plane of the sample by optically pumping dye doped solution with a stripe shaped spot (a 64$\mu$m wide and 1.8mm long stripe is pumped by a Nd:YAG laser with pulse duration of 9 ns, 120 mJ maximum intensity per pulse).The produced ASE results in an unpolarised beam characterized by a divergence of 2.2$^\circ$ (a picture is reported in figure \ref{fig:setup}a). Its spectral content is a line narrowed ($\sim$ 20 nm broad) fluorescence emission centred at 590 nm.

The beam is aligned to impinge on the drop edge by properly orienting the pumping stripe by turning a cylindrical lens (sketch in panel \ref{fig:setup}c). This approach presents several advantage versus direct illumination: I) a perfect alignment of the probe beam to the plane of the sample, II) a homogeneous illumination of thin sample side III) the reduction of artifacts because no optical interfaces are present between the diffusive liquid and the light source. Moreover contributions from pumping light has been eliminated by a colored filter.

\begin{figure}[t]
\includegraphics[width=8 cm]{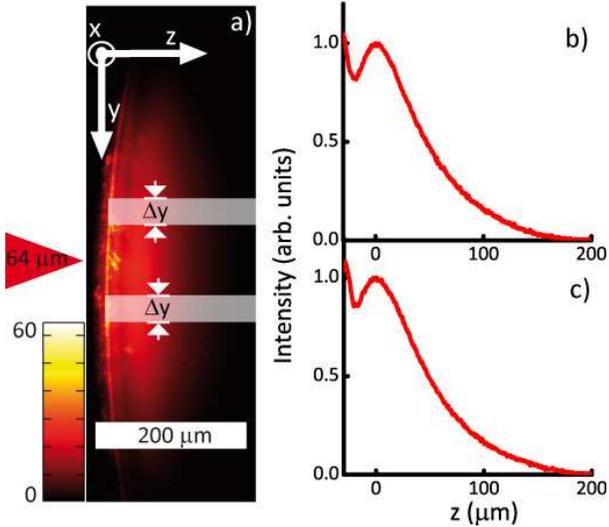} % Here is how to import EPS art
\caption{\label{fig:MeasureDL10} (Color Online) a) Color coded image for the intensity of laterally escaping light. Graphs b) and c) show $I(z)$ relative respectively to the upper and lower shaded area of a).}
\end{figure}

By using an enlarging telescope (x10 magnification, 0.4 numerical aperture), we project an image of sample top, i.e., one of the coverslips, on a CCD camera (Pixelink, pixel size 3.2$\mu$m). This image contains the information of how much light is being scattered out of the sample at each point (see Fig. \ref{fig:MeasureDL10}a).

Panels \ref{fig:MeasureDL10}b and \ref{fig:MeasureDL10}c  report the intensity $I$ averaged along the $y$ direction, over the range $\Delta y$=30$\mu m$ indicated in the figure versus the coordinate $z$ from the two shaded area of figure \ref{fig:MeasureDL10}a. The two graphs are very similar because $I(z)$ is nearly independent of the $y$ coordinates at which it is measured if the considered areas lie close to the light input position. $I(z)$  presents similar features in all samples we measured: a first maximum due to the scattering from the physical air liquid interface and a second (smoother) maximum due to diffusing photons followed by a monotonic decrease.

In samples much larger than $\ell$ the diffuse pattern depends on the distance between the light source and the image plane \cite{Johnson:08}. In  samples in which the thickness is comparable with transport mean free path instead, the probability for a photon to be ejected at each scattering event, that depends on the ratio between $\ell$ and $S$, introduces losses that affects the shape of $I(z)$. Monte Carlo simulations\cite{PhysRevLett.80.5321}, that allows to track photons from the source to the detector have been performed to confirm this picture.

\begin{figure}[h]
\includegraphics[width=8 cm]{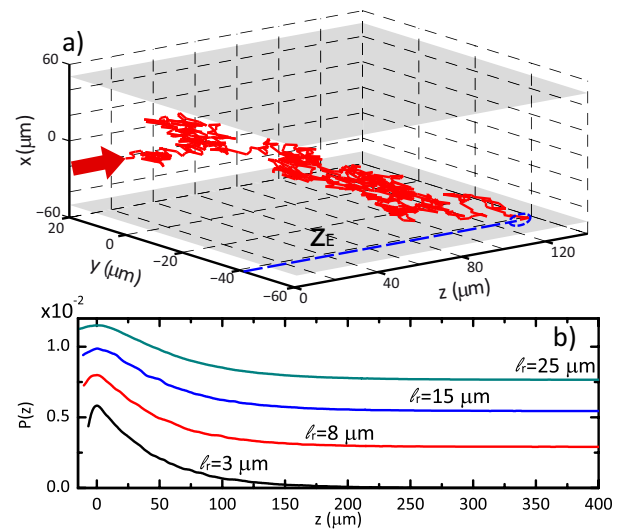} % Here is how to import EPS art
\caption{\label{fig:Sims}(Color Online) Panel a) shows the trajectory of a single photon particle with a random walk step of R=3 $\mu m$. On b) $P(z)$ for different values of $\ell_r$ is reported. Graphs are translated vertically an arbitrary amount for clarity.}
\end{figure}

In our numerical model random walking particles start from a punctual source (located at coordinates [0 0 0] and one mean free path inside the diffusive material \cite{PhysRevLett.56.1471}) and travel performing a isotropic random walk with step size equal to $\ell_r$ until they encounter sample boundary (thus escaping from top of the cell) at a certain $z$. The simulation is stopped if either the particle reaches one of the lateral edges (located at $x$=50 and $x$=-50 $\mu$m) or the physical boundary of the input edge (located at $z=-\ell_r$)\cite{PhysRevB.46.14475} (see  figure \ref{fig:Sims}a).  The coordinate $z_E$ at which the random walking particle leaves the sample is recorded. A sketch of a single simulation (consisting on a number of random walk steps of the order of 10$^2$) reporting the trajectory performed by a single photon is represented in figure 3a. By repeating the simulation, we obtained (in about one day's time of calculation in a standard personal computer) $2\cdot10^6$ values of $z_E$ that, organized in a normalized histogram, represent an estimate of the probability $P(z,\ell_r)$ of reaching a certain depth $z$ inside de sample.

Some of the curves $P(z,\ell_r)$  that have been calculated for $\ell_r$ in the range 1-40 $\mu$m are shown in figure \ref{fig:Sims}b. Each simulation included the effect of reflectance at interfaces that has been calculated following reference \cite{Pine_PRA} and has been taken into account by reducing the escaping probability of photons. Although it may be easily taken into account\cite{P_sheng}, absorption has been neglected because water and latex are transparent at the considered wavelength. Trying with different kind of sources (gaussian instead of punctual), or with more elaborated random walk steps, produced differences lower than the simulation noise.

\begin{figure}[h]
\includegraphics[width=8 cm]{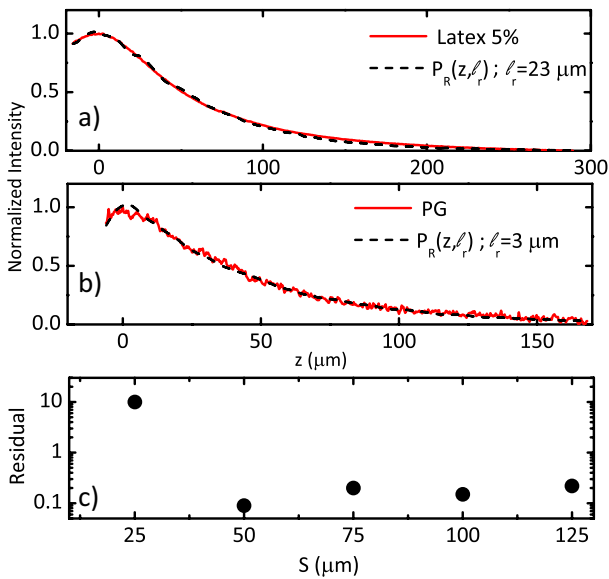} % Here is how to import EPS art
\caption{\label{fig:Exps_Sims}(Color Online) Measurements and best fits for sample "Latex 5\%" a) and for sample "PG" b). c) reports fit residuals as a function of the spacers size $S$ for the sample "Latex 10\%".}
\end{figure}

On the other hand the numerical calculation of $P(z,\ell_r)$ for a given $\ell_r$ is performed once and used to extrapolate transport mean free path for the considered samples simply by rescaling its maximum to coincide with the experimental one ($P_R(z,\ell_r)=\alpha P(z,\ell_r)$). A good estimate of $\ell$  is the $\ell_r$ that produces a curve $P_R(z,\ell_r)$ that best fits the $I(z)$ of the considered sample. Figures \ref{fig:Exps_Sims}a and \ref{fig:Exps_Sims}b show the best fitting $P_R(z,\ell_r)$ together with $I(z)$ for two different systems: a dispersion of latex beads with concentration of 5\% and  a photonic glass (that is a disordered matrix of latex beads blocked in fixed position) obtained by the evaporation of the water content from a latex dispersion as described in reference \cite{PhysRevA.78.023823}.

Agreement between our experimental protocol and standard techniques\cite{PhysRevA.78.023823, JJAP.39.146} is confirmed by table \ref{table1} in which our measurements are reported together with results from literature. Samples consist in water dispersion of latex beads (520 nm in size) with volume fraction 5\% and 10\%, respectively indicated with "Latex 5\%" and "Latex 10\%". Sample "PG" is the photonic glass. Reported values and errors result from the statics of the measurements on 5 samples for each kind.

\begin{table}[h]
%\begin{ruledtabular}
\begin{tabular}{ccc}
Sample & $\ell(\mu m)$ Measured &  $\ell (\mu m)$ From Refs.\\ \hline
Latex 5\% & $19\pm4$ & $\sim21^\dag$ \\
Latex 10\% & $8\pm2$ & $\sim12^\dag$ \\
PG  &   $3\pm1$ &  $\sim$2-3$^\sharp$\\ \hline
\end{tabular}
\caption{\label{table1} Measured $\ell$ compared with literature. $^\dag$ are from reference \cite{JJAP.39.146} while $^\sharp$ is relative to  \cite{PhysRevA.78.023823}. }%\end{ruledtabular}
\end{table}

Graph \ref{fig:Exps_Sims}c reports residual of the fitting process (defined as the absolute value of the area difference between the experimental and numerical normalized curves) obtained measuring sample "Latex 10\%"  versus the  spacers thickness $S$. A good agreement is evident from $S\geq 50 \mu m\sim 6 \times \ell$, conversely for lower values of $S$ the contribution from single scattering photons is so strong that the shape of $I(z)$ is no longer in agreement with diffusion approximations\cite{PhysRevLett.64.2647}.

In conclusion by collecting intensity exiting laterally in a thin diffusive system shined with light generated by in-plane ASE, and by comparing our measurements with predictions from numerical random walk simulation, we are able measure the mean free path. We demonstrated how our results are in agreement with measurements from standard techniques, and that by using our approach samples with a thickness up to 6 times the mean free path may be characterized.  Our measurement protocol allows a systematic study of light diffusion in intrinsically thin samples like those fabricated by nanolithography or self assembly and optically thin biological tissue.

The work was supported by EU FP7 NoE Nanophotonics4Enery Grant No 248855; The work was supported by the Spanish MICINN CSD2007-0046 (Nanolight.es), MAT2009-07841 (GLUSFA) and Comunidad de Madrid S2009/MAT-1756 (PHAMA).

\end{document}